\documentclass{emulateapj} 
\usepackage{graphicx,amssymb,enumerate}
\usepackage{natbib}

\newcommand{\kms}{km s$^{-1}$}

\shorttitle{ALFALFA Velocity Width Function}
\shortauthors{Papastergis et al.}

\begin{document}

\title{The Velocity Width Function of galaxies from the 40\% ALFALFA survey: shedding light on the cold dark matter overabundance problem}

\author {Emmanouil Papastergis\altaffilmark{1}, Ann M. Martin\altaffilmark{1}, Riccardo Giovanelli\altaffilmark{1,2}, Martha P. Haynes\altaffilmark{1,2}}
\altaffiltext{1}{Center for Radiophysics and Space Research, Space Sciences Building,
Cornell University, Ithaca, NY 14853. {\textit{e-mail:}} papastergis@astro.cornell.edu, amartin@astro.cornell.edu,riccardo@astro.cornell.edu, haynes@astro.cornell.edu}
\altaffiltext{2}{National Astronomy and Ionosphere Center, Cornell University,
Ithaca, NY 14853. The National Astronomy and Ionosphere Center is operated
by Cornell University under a cooperative agreement with the National Science
Foundation.}

\begin{abstract}
The ongoing Arecibo Legacy Fast ALFA (ALFALFA) survey is a wide-area, extragalactic HI-line survey conducted at the Arecibo Observatory. Sources have so far been extracted over $\sim$ 3000~deg$^2$ of sky (40\% of its final area), resulting in the largest HI-selected sample to date. We measure the space density of HI-bearing galaxies as a function of their observed velocity width (uncorrected for inclination) down to $w = 20$ \kms, a factor of 2 lower than the previous generation HI Parkes All-Sky Survey. We confirm previous results that indicate a substantial discrepancy between the observational distribution and the theoretical one expected in a cold dark matter (CDM) universe, at low widths. In particular, a comparison with synthetic galaxy samples populating state-of-the-art CDM simulations imply a factor of $\sim 8$ difference in the abundance of galaxies with $w = 50$ \kms \ (increasing to a factor of $\sim 100$ when extrapolated to the ALFALFA limit of $w = 20$ \kms). We furthermore identify possible solutions, including a keV warm dark matter scenario and the fact that HI disks in low mass galaxies are usually not extended enough to probe the full amplitude of the galactic rotation curve. In this latter case, we can statistically infer the relationship between the measured HI rotational velocity of a galaxy and the mass of its host CDM halo. Observational verification of the presented relationship at low velocities would provide an important test of the validity of the established dark matter model.

\end{abstract}

\section{Introduction}
\label{sec:introduction}

The current ``standard''  $\Lambda$CDM cosmological model has been extremely successful at reproducing the bulk of the observed properties of our universe on large scales \citep{2011ApJS..192...18K}. However, given the current lack of a firm theoretical understanding of dark energy and the lack of a direct or indirect detection of the dark matter (DM) particle \citep{2009PhRvL.102a1301A, 2008PhRvL.100b1303A, 2008PhRvL.101i1301A, 2010PhRvL.104i1302A, 2010ApJ...712..147A, 2009Natur.458..607A}, it is important to test in detail the astrophysical implications of the established cosmological paradigm.

One of the most interesting consequences of assuming a cold dark matter (CDM) model is that substructure forms first on small scales, resulting in a present-day universe populated by a multitude of low-mass halos. More formally, the mass distribution of DM halos is described by the DM mass function (MF), which is defined as the number density of halos as a function of their virial mass; it can be analytically predicted \citep{1974ApJ...187..425P, 2002MNRAS.329...61S} that the MF displays a power-law behavior at low halo masses, $n \propto M^\alpha$, with a relatively steep exponent of $\alpha \approx -1.9$ in the standard $\Lambda$CDM context. This analytical expectation, confirmed to great accuracy by N-body simulations of structure formation \citep{2006ApJ...646..881W, 2009MNRAS.398.1150B, 2010arXiv1002.3660K}, leads to the prediction of a large number of low mass halos for every Milky Way-sized (MW-sized) halo found in the present epoch.     

This firmly established theoretical result has led to a number of observational challenges, such as the ``missing satellites problem'' \citep{1999ApJ...522...82K, 1999ApJ...524L..19M, 2007ApJ...667..859D, 2007ApJ...669..676S, 2007ApJ...670..313S}, the ``void phenomenon'' \citep{2001ApJ...557..495P, 2009ApJ...691..633T}, as well as the discrepancy between the sizes of mini-voids observed in the local universe and those produced in CDM simulations \citep{2009MNRAS.395.1915T}. Additional concerns, again closely related to the distribution of halo masses predicted by CDM, are raised by the flatness of the galactic luminosity function \citep[LF,][]{2005ApJ...631..208B, 2009MNRAS.399.1106M}, HI mass function \citep[HIMF,][]{2010ApJ...723.1359M} and galactic stellar mass function \citep[GSMF,][]{2008MNRAS.388..945B, 2009MNRAS.398.2177L} at their faint/low-mass end. These observational distributions display power-laws with $\alpha \approx -1.3$, much shallower than expected from the combination of a CDM universe plus a naive linear relationship between halo mass and luminosity/baryonic mass. Despite their apparent diversity, all statements described above are just different aspects of the same fundamental issue: CDM structure formation predicts large numbers of low mass halos, seemingly in contradiction with the relative paucity of visible low-mass galaxies. Hereafter, we refer to this discrepancy as the \textit{CDM overabundance problem}\footnote{This statement does not aim at including a second class of potential observational challenges to CDM, related to the density profile of halos in their central regions (known as the ``cusp versus core'' problem).}.

The main caveat regards the proper interpretation of these observational results. All phenomena mentioned so far rely on the measurement of quantities indirectly related to the mass of the hosting DM halo (e.g. luminosity or HI/stellar mass) and, as a result, do not provide a direct means of comparing the MF expected for CDM with the MF realized in nature. 
In fact, a number of environmental and feedback effects (see \S\ref{subsec:solutions}) are expected to affect the baryonic content of halos, with low mass ones being the most impacted.

Ideally, one would need a large sample of galaxies with directly measured dynamical masses (e.g. through lensing or satellite kinematics), extending all the way to the low mass regime. Unfortunately, current datasets are restricted to relatively massive galaxies. The best practical alternative would consist of a rich sample of resolved HI-interferometric rotation curves of  galaxies, spanning a wide range in dynamical mass. Atomic hydrogen is usually the most spatially extended baryonic component in a galaxy, and therefore the best tracer of the rotation curve at large galactic radii. Such a sample could be used to determine the space density of galaxies as a function of their measured maximum rotational velocity, $v_{rot}$. This observational statistic, which is referred to as the velocity function (VF) of galaxies, is more directly related to the halo dynamics than statistics based on luminosity/baryonic mass and has a largely different set of systematic issues. However, current datasets are very limited, mostly because HI interferometry is extremely time consuming (especially for low HI-mass targets).

A more economical approach is to rely on wide-area, single-dish 21 cm surveys. Thanks to their intrinsic spectroscopic nature, HI surveys automatically obtain the spectral HI-line profile of every detected source. The velocity width of each detected galaxy, $w$, can thus be readily extracted, and the associated dataset can be used to measure the velocity width function (WF) of galaxies.  One can furthermore apply inclination corrections to the measured widths in order to retrieve intrinsic rotational velocities ($v_{rot}$), and then estimate the galactic VF. Correcting for inclination requires however the use of external datasets, usually optical/NIR photometric surveys.

So far, the most accurate WF and VF for late-type galaxies have been based on 4315 and 2646 HI-selected galaxies respectively, detected by the HIPASS survey \citep[hereafter Zw10]{2010MNRAS.403.1969Z}. Their measurement of the VF extends over the velocity range $30$ \kms $< v_{rot} < 300$ \kms, and suggests a dramatic departure from the CDM expectation at low velocities ($v_{rot} \lesssim 100$ \kms). Recent determinations of the VF for massive early-type galaxies (which are mostly absent from HI-selected samples) have been obtained using the Sloan Digital Sky Survey (SDSS) and Two-degree Field Galaxy Redshift Survey (2dFGRS) datasets by \cite{2007ApJ...658..884C} and \cite{2010MNRAS.402.2031C}. Both the late-type and early-type distributions are needed in order to derive the ``total'' galactic VF , since massive early-type galaxies are the dominant population at high velocities ($250$ \kms $\lesssim v_{rot} \lesssim 450$ \kms) while late-types dominate the counts at lower velocities ($v_{rot} \lesssim 250$ \kms).

In this paper we present the Arecibo Legacy Fast ALFA (ALFALFA) measurement of the velocity width function of HI-bearing galaxies. The decision not to correct the measured widths for inclination is intentional, as the WF maintains all the advantages of the VF as a probe of the halo mass distribution, while featuring a number of observational advantages over the latter (see Sec. \ref{sec:wf} for more details). The ALFALFA WF is based on 10,744 HI-selected galaxies (a more than twofold increase over previous datasets) and extends to widths as low as $w = 20$ \kms. 

This paper is organized as follows: in Section \ref{sec:dataset} we present the ALFALFA survey and the associated dataset; in Section \ref{sec:wf} we discuss the observational advantages of the WF with respect to the inclination-corrected VF and present the ALFALFA measurement of the WF for HI-bearing galaxies; in Section \ref{sec:biases} we address possible observational biases on the determination of the ALFALFA WF; in Section \ref{sec:theory} we compare the ALFALFA measurement with the expectations in a CDM universe, and describe the possible solutions to the observed discrepancy at low widths; in Section \ref{sec:relation} we derive the relation between $v_{rot}$ (measured observationally) and $v_{halo}$ (calculated from N-body simulations), that would be needed to reconcile the velocity distributions of CDM halos and observed galaxies. We conclude with Section \ref{sec:conclusions} by summarizing our results. 

Throughout this paper we use a Hubble constant of $H_0 = 70$ \kms; $h_{70}$ refers to the Hubble constant in units of 70 km s$^{-1}$ Mpc$^{-1}$, while $h$ refers to the Hubble constant in units of 100 km s$^{-1}$ Mpc$^{-1}$.

\section{ALFALFA dataset}
\label{sec:dataset}

\subsection{The survey}

The ongoing ALFALFA survey is a wide-area, blind 21 cm emission-line survey that takes advantage of the increased survey speed offered by the 7-feed Arecibo L-band Feed Array (ALFA) receiver at the Arecibo Observatory. The ALFALFA data are acquired in a minimally invasive drift-scan mode in two passes, ideally separated by several months in order to enable the discrimination between narrow-band radio frequency interference (RFI) and small spectral width cosmic signals. When complete, the survey will have detected $>$30,000 galaxies over an area of $\sim$~7000 deg$^2$ of sky  out to $cz \approx$ 18,000 \kms. The ALFALFA survey is more sensitive than the previous generation HIPASS survey \citep{2004MNRAS.350.1195M, 2004MNRAS.350.1210Z}, with a 5$\sigma$ detection limit of 0.72~Jy~\kms \ for a source with a profile width of 200 \kms \ as compared to a 5$\sigma$ sensitivity of 5.6~Jy~\kms \ for the same source in HIPASS \citep{2005AJ....130.2613G}. In addition to greater sensitivity, ALFALFA has a finer velocity resolution (11.2 \kms \ versus 26.4 \kms \ for smoothed data) and better angular resolution ($ 3.6^\prime$ vs. 13$^\prime$ FWHM), resulting in a more accurate identification of optical counterparts.

\subsection{The sample}
\label{subsec:sample}

ALFALFA catalogs have so far been extracted \citep{2007AJ....133.2569G, 2008AJ....135..588S, 2008AJ....136..713K, 2009AJ....138..338S, 2009ApJS..183..214M,2011arXiv1109.0027H} for a total area of 2934 deg$^2$. The current ALFALFA footprint consists of four distinct regions: two in the northern Galactic hemisphere, hereafter referred to as the Virgo direction region (VdR: $07^h30^m < \alpha < 16^h30^m$, $4^\circ < \delta < 16^\circ$ and $24^\circ < \delta < 28^\circ$), and two in the southern Galactic hemisphere, hereafter referred to as the anti-Virgo direction region (aVdR: $22^h < \alpha < 03^h, \; 14^\circ < \delta < 16^\circ$ and $24^\circ < \delta <32^\circ$). From this primary dataset we only select  extragalactic objects detected at high significance ($ S/N > 6.5$, designated Code 1), and we further restrict ourselves to the redshift range $cz \leqslant 15,000$ \kms, beyond which interference from the nearby San Juan airport causes a significant drop of the ALFALFA detection efficiency. This final sample, corresponding to $\sim 40\%$ of the ALFALFA survey area (hereafter $\alpha$.40 sample), contains a total of 11,086 galaxies.

\begin{figure}[p]
\includegraphics[scale=0.65]{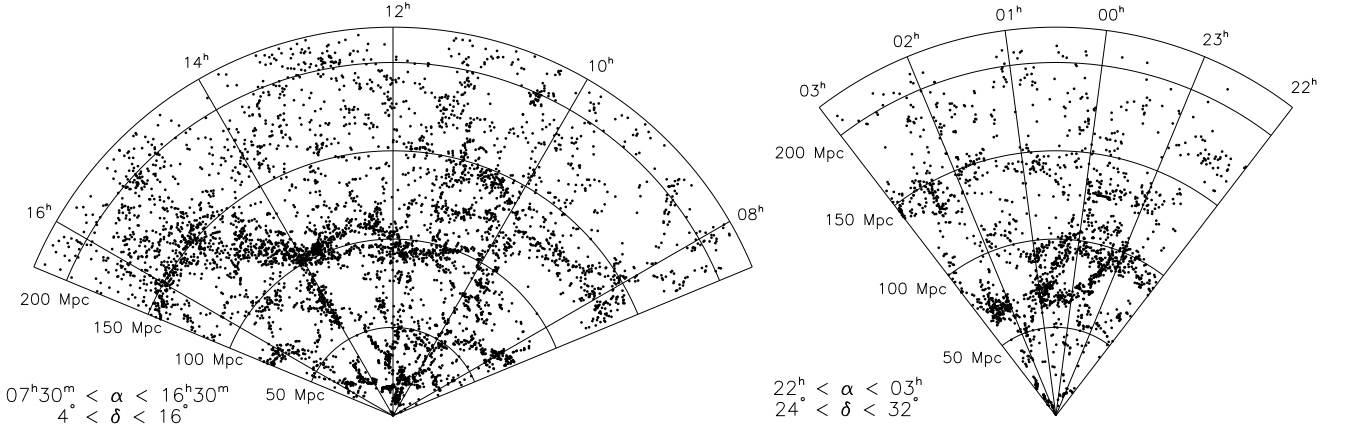}
\caption{Spatial distribution of 5868 sources in the Virgo direction region (VdR, \textit{left panel}) and 2055 sources in the anti-Virgo direction region (aVdR, \textit{right panel}). The Virgo Cluster and the ``Great Wall'' are the most conspicuous structures in the VdR (located at a distance of  $\approx 17$ Mpc and $\approx 100$ Mpc respectively). In the aVdR, the Pisces-Perseus Supercluster (clearly visible at $\approx 70$ Mpc) as well as the void in its foreground dominate the large-scale structure. Distances are assigned through a combination of a flow model for the nearby Universe and Hubble distances for more distant galaxies (see \S\ref{subsec:sample}).}
\label{fig:coneplots}
\end{figure}

Figure \ref{fig:coneplots} shows the spatial distribution of the $\alpha$.40 sources in the Virgo and anti-Virgo directions respectively, and puts in evidence the complex large-scale structure present in both volumes. Density fluctuations in the survey volume can be the dominant source of statistical uncertainty in surveys like ALFALFA, where the sample size ensures small counting errors. Our statistical estimator, described in \S\ref{subsec:aawf}, in chosen to minimize this structure-induced bias.

\begin{figure}[p]
\includegraphics[scale=0.7]{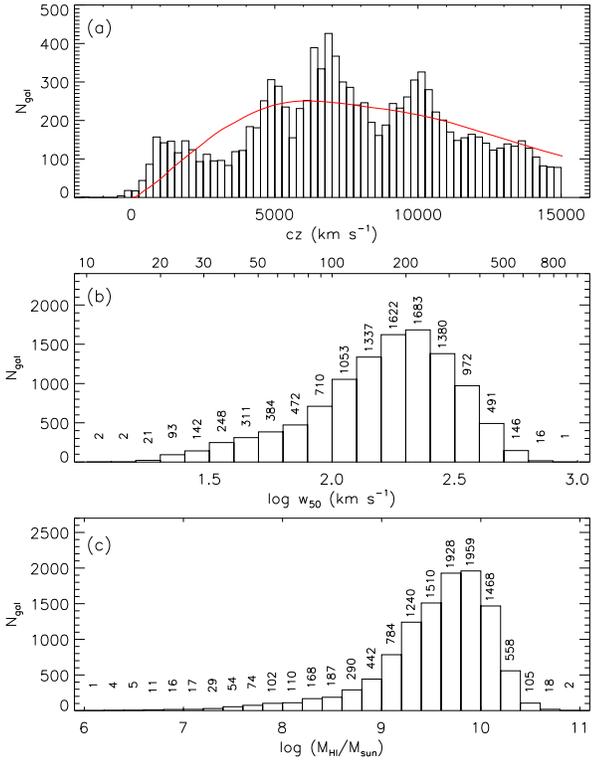}
\caption{\textit{Properties of the $\alpha$.40 sample}: Histogram (a) represents the distribution in heliocentric velocity ($v_\odot$), while the red solid line represents the distribution expected in a homogeneous universe according to the selection function of the survey; the complex large-scale structure in the survey volume is apparent. Histogram (b) represents the distribution of velocity width ($w_{50}$); note the large number of very low-width galaxies ($w_{50} < 30$ \kms) detected. Histogram (c) represents the distribution of galaxy HI mass ($M_{HI}$); again note the detections at very low HI mass ($M_{HI} < 10^{8} \: M_\odot$).}
\label{fig:histo}
\end{figure}

Figure \ref{fig:histo} displays some statistical properties of the $\alpha$.40 sample. Histograms (a) and (b) represent the distribution of heliocentric velocity, $v_\odot$, and of signal profile width, $w_{50}$, which are both directly measured quantities \citep{2007AJ....133.2569G, 2007AJ....133.2087S}. The signal profile width is measured at the 50\% flux level of each of the two peaks of the typical double-horned HI profile (or at 50\% of the single peak flux, for single-peaked profiles). The value of $w_{50}$ reported in the ALFALFA catalogs is further corrected for instrumental broadening. Histogram (c) displays the distribution of galaxy HI mass, $M_{HI}$, which is a distance dependent (and hence derived) quantity. Unlike previous HI surveys, we assign distances to nearby galaxies through a peculiar velocity flow model (Masters~2005) and use Hubble flow distances only for galaxies with $cz > 6000$ \kms \ (see \S\ref{subsec:eddington} for a detailed discussion on the impact of distance uncertainties on ALFALFA results).

\begin{figure}[p]
\includegraphics[scale=0.6]{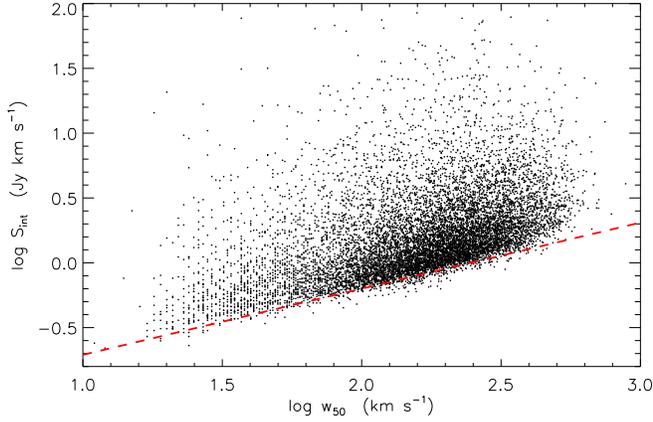}
\caption{Distribution of the $\alpha$.40 sources in the velocity width vs. integrated flux ($w_{50} - S_{int}$) plane. The dashed red line is the survey completeness limit adopted in this work $(S_{int,lim}/ 1 \: \mathrm{Jy} \: \mathrm{km}\: \mathrm{s}^{-1}) =0.06 \; (w_{50}/1 \: \mathrm{km}\: \mathrm{s}^{-1})^{0.51}$, which follows very closely the theoretically expected $S_{int,lim} \propto w^{1/2}$.}
\label{fig:flux}
\end{figure}

Figure \ref{fig:flux} displays the distribution of $\alpha$.40 sources in the velocity width ($w_{50}$)  versus integrated-flux ($S_{int}$) plane. As expected, the detection limit of the survey is a function of signal profile width and correctly scales as $S_{int,lim} \sim w_{50}^{1/2}$. Due to the large density of sources near the detection limit, we evaluate the completeness limit of the survey (red dashed line in Figure \ref{fig:flux}) based on the actual data rather than on simulations using synthetic sources.

\section{The Velocity Width Function}
\label{sec:wf}

We obtain rest-frame galaxy velocity widths, $w$, by correcting the cataloged profile widths ($w_{50}$) for Doppler broadening. It is customary to apply additional inclination corrections to $w$, in order to recover intrinsic rotational velocities, $v_{rot}$. However, since most extragalactic sources are unresolved at centimeter wavelengths, such corrections rely on external datasets (usually optical or NIR photometric surveys) for the determination of galaxy inclinations. Here, we choose to make no further corrections to $w$ and measure the velocity width function (WF) of galaxies, denoted by $\phi(w)$. 

Even though the WF does not directly represent the distribution of any fundamental galaxy property, it is observationally superior to the (inclination-corrected) VF. In particular, it is free of the restrictions and systematics that arise from cross-matching HI and optical catalogs and correcting for galaxy inclination. For example, the HIPASS primary sample contains 4315 sources of which only 2646 have unambiguous optical counterparts \citep{2005MNRAS.361...34D}. Another 30\% of the sources in this restricted subsample have low inclination values ($i < 45^\circ$), and are thus subject to large inclination-correction errors. As a result, only $\approx 43 \%$ of the galaxies in the HIPASS primary sample were used for their determination of the VF. Furthermore, obtaining accurate estimates of the true orientation of irregularly-shaped dwarf galaxies is challenging, and the process may introduce biases in the measurement of the low-velocity end of the VF. 

Nonetheless, a measurement of the galactic WF would not be useful if it did not provide an accurate means of comparing the outcome of N-body simulations with the observed universe. Fortunately, it is relatively straightforward to project a given theoretical rotational velocity distribution and transform it into its corresponding width distribution (see \S\ref{subsec:overabundance}). We conclude that the WF should be regarded as the \textit{prime observational distribution} for single-dish HI surveys, against which to compare theoretical expectations.

\subsection{The ALFALFA Velocity Width Function}
\label{subsec:aawf}

\begin{figure}[p]
\includegraphics[scale=0.6]{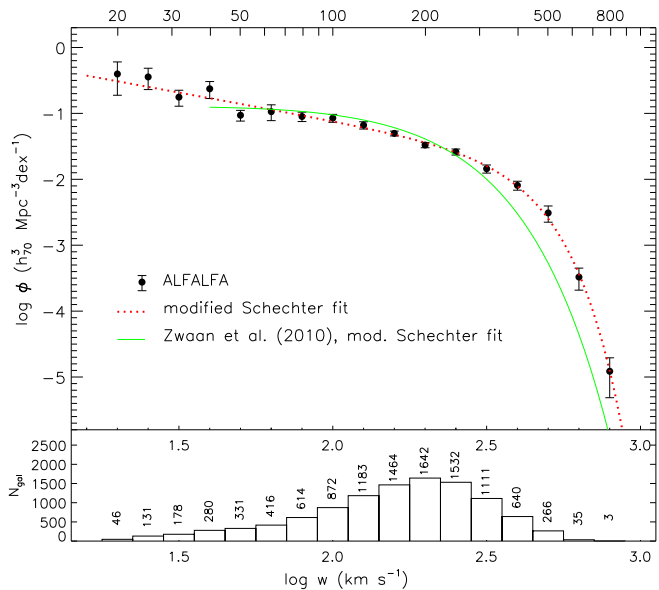}
\caption{\textit{The ALFALFA velocity width function (WF)}: datapoints represent the space density of HI-bearing galaxies as a function of velocity width (corrected for Doppler and instrumental broadening, but uncorrected for inclination), as inferred from 10,744 galaxies detected by the 40\% ALFALFA survey. The errors are $1\sigma$ Poisson errors due to galaxy counts in individual width bins. The red dotted line corresponds to a modified Schechter fit to the ALFALFA WF (see \S\ref{subsec:aawf}). The green solid line represents the fit to the HIPASS WF based on 4315 galaxies \citep{2010MNRAS.403.1969Z}, over its measured range.}
\label{fig:wf}
\end{figure}

In Figure \ref{fig:wf} we present the ALFALFA width function, based on 10,744 galaxies drawn from the $\alpha$.40 sample. For the calculation of the WF we restrict ourselves to $\alpha$.40 galaxies which are positioned in the portion of the flux-width plane where the ALFALFA survey is complete (i.e. above the red dashed line in Figure \ref{fig:flux}) and have profile widths broader than $w_{50} \gtrsim 18$ \kms. This cut results in the elimination of  $\approx 330$ galaxies from the calculation. An additional 13 very nearby sources are eliminated, for which the flow model assigned distances are subject to large uncertainty. 

The WF is calculated in logarithmic width bins, according to the $\Sigma \, 1/V_{eff}$ method \citep{2005MNRAS.359L..30Z}. The $\Sigma \, 1/V_{eff}$ method is a non-parametric maximum likelihood method and, as such, it is insensitive to the presence of large-scale structure in the survey volume. As its name suggests, it closely resembles the traditional $\Sigma \,1/V_{max}$ method \citep{1968ApJ...151..393S} and consists of summing the number of detections in each width bin, weighted by the inverse of the ``effective'' volume available to each source. More precisely, the space density of galaxies belonging to width bin $k$ ($k=1,2,...,N_w$) is

\begin{equation}  
\label{eqn:veff}
\phi_k = \sum_i \frac{1}{V_{eff,i}} \; \; \; \; \mathrm{for}\: \mathrm{all} \: \mathrm{galaxies} \: i \: \mathrm{in}\: \mathrm{width} \: \mathrm{bin} \: k \; .
\end{equation}  

\noindent
In the case of a spatially homogeneous survey volume, $V_{eff,i}$ would coincide with $V_{max,i}$, the latter defined as the volume within which galaxy $i$ could be placed and still be detectable by the survey. On the other hand, if the survey volume displays significant density variations, $V_{eff,i}$ takes into account the relative density of the volume available to galaxy $i$ with respect to the mean density of the total survey volume. As with all density-independent estimators, the overall normalization is lost, and has to be calculated afterwards. The normalization is fixed by matching the integral of the distribution to the average number density of galaxies in the survey volume (see \citealp[Appendix B.1]{2010ApJ...723.1359M} for more details).

Due to its spectral resolution and sensitivity, ALFALFA can push the low-width limit of the WF to $w \approx 20$~\kms, a factor of 2 lower than the HIPASS survey. Over the full measured range (20 \kms $< w < 800$ \kms) the ALFALFA WF is very well described by a modified Schechter function of the form\footnote{The parameterization here is equivalent to the parameterization $\phi(w) \: dw = \phi_\ast \: (w/w_\ast)^\alpha \: \exp{-(w/w_\ast)^\beta} (\beta/\Gamma(\alpha/\beta)) \: dw/w$ presented by other authors, except for the normalization factor $\beta/\Gamma(\alpha/\beta)$.} 

\begin{equation}
\phi(w) = \frac{dn}{d\log w} = \ln(10) \: \phi_\ast \left(\frac{w}{w_\ast}\right)^\alpha e^{-(\frac{w}{w_\ast})^\beta} \; \;.
\label{eqn:modschechter}
\end{equation} 

\noindent
The least squares parameters\footnote{The least squares parameters and their statistical errors were determined by the MPFITFUN procedure, written in the IDL programming language.} are $\phi_\ast = 0.011 \, \pm \, 0.002 \;\; h_{70}^3 \; $Mpc$^{-3}$dex$^{-1}$, $\log w_\ast = 2.58 \, \pm \, 0.03$, $\alpha = -0.85 \, \pm \, 0.10$ and $\beta = 2.7 \, \pm \, 0.3$ (uncertainties are statistical $1\sigma$ errors due to Poisson errors on the individual bin values). Note, however, that the final sample contains 163 sources that lack a confidently identified optical counterpart. Some of these sources correspond to tidal debris from nearby interacting galaxies and may not be hosted by individual DM halos. Excluding these galaxies from the WF calculation leads to a somewhat shallower narrow-end slope of $\alpha = -0.68 \, \pm \, 0.11$.

\begin{figure}[p]
\includegraphics[scale=0.6]{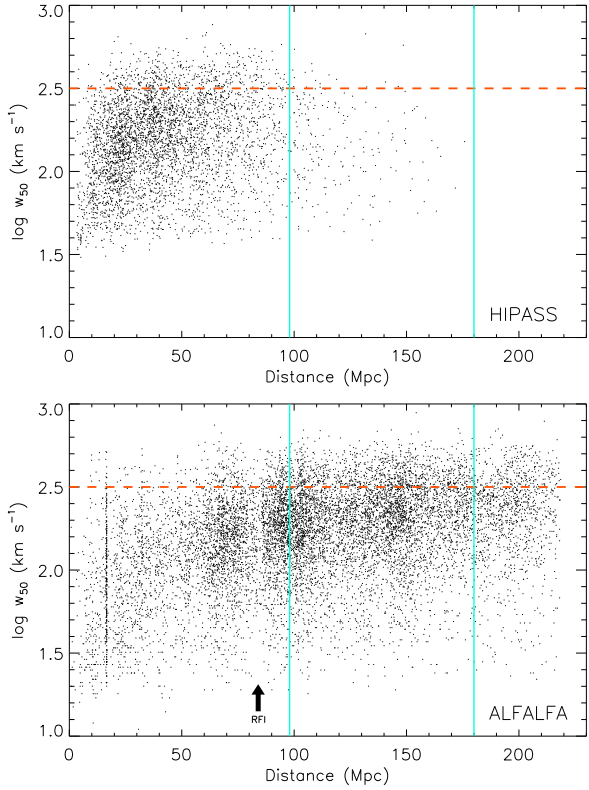}
\caption{``velocity width Spanhauer'' diagrams for ALFALFA (\textit{bottom}) and HIPASS (\textit{top}) on the same scale. The region above the horizontal orange line marks the range over which the two width functions disagree. Despite the fact that the nominal value of the HIPASS volume is a factor of $\sim$5 larger than the $\alpha$.40 volume, ALFALFA detects more very broad profile galaxies. This is due to the limited sensitivity of HIPASS, which leads to a ``thinning'' of detections beyond $\approx$ 100 Mpc and out to the survey boundary (area enclosed by vertical cyan lines).}
\label{fig:spanhauer}
\end{figure}

ALFALFA finds significantly more high-width galaxies than HIPASS (a factor of $\sim 3$ at $w \approx 400$ \kms, growing to a factor of $\sim 10$ at $w \approx 800$ \kms), which is also evident from the marked difference in the value of the position of the ``knee'' of the WF for the two surveys ($\log w_\ast  = 2.58 \, \pm \, 0.03$ for ALFALFA versus $\log w_\ast = 2.21 \, \pm \, 0.10$ for HIPASS\footnote{No errors are reported for the published fit parameters to the HIPASS WF. In order to compare with ALFALFA, we derive errors by performing a least squares fit to the HIPASS WF datapoints.}, in disagreement at the $> 3\sigma$ level). Despite the fact that the nominal HIPASS volume is a factor of $\sim 5$ larger than the $\alpha$.40 volume, ALFALFA is able to find more high-width galaxies thanks to its better sensitivity (see Figure \ref{fig:spanhauer}). The same effect can be seen in the HIMFs published by the two surveys, with ALFALFA \citep{2010ApJ...723.1359M} finding a factor of a few more of the highest HI-mass galaxies compared to HIPASS \citep{2005MNRAS.359L..30Z}. 

On the low-width end, ALFALFA finds a rising slope ($\alpha < 0$) which is, however, by no means steep enough to match the CDM prediction (see Sec. \ref{sec:theory}). Despite the vastly different value for the narrow-end slope reported by the two surveys ($\alpha = 0.10 \, \pm \, 0.39$ for HIPASS versus $\alpha  = -0.85 \, \pm \, 0.10$ for ALFALFA) the HIPASS and ALFALFA datapoints are consistent in the width range $40$~\kms \ $ \lesssim w \lesssim 200$~\kms. The HIPASS $\alpha$ parameter is not well constrained, as their WF does not extend to low enough widths and suffers from considerable counting error in the low-width bins.

\section{Biases}  
\label{sec:biases}

\subsection{Measurement errors on $w_{50}$}
\label{subsec:widtherror}
 
Measurement errors on $w_{50}$ can shift galaxies among width bins, altering the bin counts and therefore the inferred space density. The $w_{50}$ value for ALFALFA sources is subject to two separate sources of error: one is statistical in nature and present for all sources, while the other is systematic and concerns only a fraction of the $\alpha$.40 sample. The former is due to the distortion of the signal profile shape by noise; the latter results from the fact that the measurement of the spectral width of a signal relies on the accurate visual identification of its spectral boundaries, which is non-trivial for a number of sources (especially those found in the vicinity of RFI). 
The final width error reported in the ALFALFA catalogs, $\Delta w_{50}$, is the sum in quadrature of the random and systematic error terms described above. Owing to the fact that all $\alpha$.40 galaxies are detected with high signal to noise and have a clean spectral profile in the vast majority of cases, the typical $\alpha$.40 width error is relatively small and its distribution well behaved. The median error is $\Delta w_{50,median} \approx 8$~\kms \ and $\sim$70\% of the sources have a fractional error of $\Delta w_{50} / w_{50} \leqslant 10$\%. 

In order to assess the effect of $\Delta w_{50}$ on the WF, we create 50 mock galaxy samples by re-assigning random widths to every galaxy $i$ in the primary ALFALFA dataset according to their individual measured width ($w_{50,i}$) and error ($\Delta w_{50,i}$). Each mock sample is subject to the same cuts as the $\alpha$.40 sample and a new realization of the WF is calculated (``1x'' set). In order to illustrate the systematic trends introduced, we also perform an additional set of WF realizations with artificially inflated width errors (twice the reported ALFALFA width errors, ``2x'' set).    

\begin{figure}[p]
\includegraphics[scale=0.6]{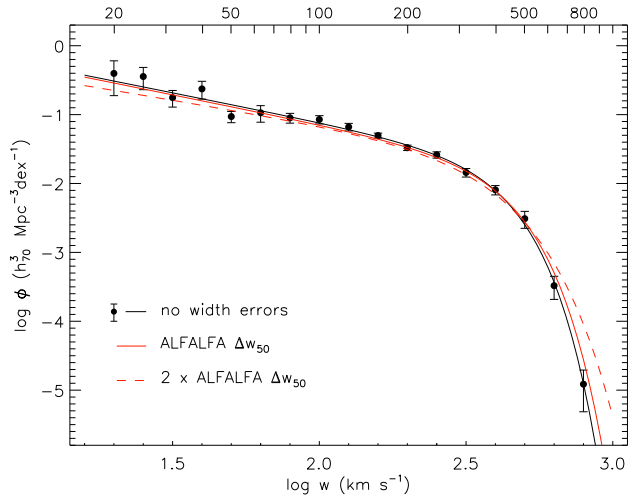}
\caption{\textit{Effect of width measurement errors on the width function}: filled circles with errorbars and the black solid line represent the ALFALFA WF and the best-fitting modified Schechter function (same as Figure \ref{fig:wf}). The red solid line corresponds to the distribution obtained by taking into account the ALFALFA measurement error on $w_{50}$. The WF remains mostly unchanged, except perhaps for a slight increase at the high width end. The red dashed line corresponds to artificially inflated width errors (twice the $\alpha$.40 errors) and is plotted in order to illustrate the general systematic trend introduced by width errors on the WF (see \S\ref{subsec:widtherror} for more details).}     
\label{fig:widtherrors}
\end{figure}

The results are shown in Figure \ref{fig:widtherrors}: overplotted to the original ALFALFA WF (datapoints and solid black line) are a modified Schechter fit to the mean WF corresponding to the 1x (red solid line) and 2x (dashed red line) realizations. Width errors at the ALFALFA error levels seem to only slightly affect the high-width end of the WF. As evidenced by the 2x run, width errors generally lead to a rise of the high-width end, due to a net ``diffusion'' of galaxies from intermediate-width bins with large number counts to high-width bins with lower number counts.

\subsection{Distance Uncertainties}
\label{subsec:eddington}

Since velocity width is a distance-independent quantity,  galaxy counts in width bins are not altered by distance errors. However, the weights ($1/V_{eff,i}$) that each galaxy contributes to its bin depend on HI-mass (see Eqn. \ref{eqn:veff} and discussion in \S\ref{subsec:aawf}), and therefore on the assumed distance. %
\citet{2004ApJ...607L.115M} have shown that  ignoring the local peculiar velocity field can lead to biased estimates of galaxy statistical distributions, especially for surveys drawing a large fraction of their sample from the Virgo direction (VdR). To avoid this bias ALFALFA uses redshift distances only for distant ($cz > 6000$ \kms) galaxies and assigns distances to nearby galaxies through a parametric flow model developed by \citet{2005PhDT.........2M}. The model includes two attractors (Virgo Cluster \& Great Attractor), a dipole component (Local Group peculiar velocity), a quadrupole component (Local Group asymmetric expansion) and a random thermal residual of $\sigma_{local} \approx 160$ \kms. Here we assume that most of the coherent motion of nearby galaxies is correctly described by the flow model, and no significant bias results from this systematic component of galaxy peculiar velocities. Contrary to intuition however, even the random component $\sigma_{local}$ can induce a systematic bias through the ``Eddington effect'' (see for example  Figure 6 in \citealp{2003AJ....125.2842Z}).

In order to asses the effect of $\sigma_{local}$ on the WF, we proceed as in \S\ref{subsec:widtherror} and create 50 mock samples by adding gaussian noise on the cataloged distance of each $\alpha$.40 galaxy. We calculate the WF corresponding to each sample realization, and use the obtained average distribution to investigate the effect of distance uncertainties on the WF. We adopt the Masters (2005) value of $\sigma_{local} \approx 160$ \kms, but we also perform simulations with double the fiducial dispersion ($\sigma_{local} \approx 320$ \kms).

\begin{figure}[p]
\includegraphics[scale=0.6]{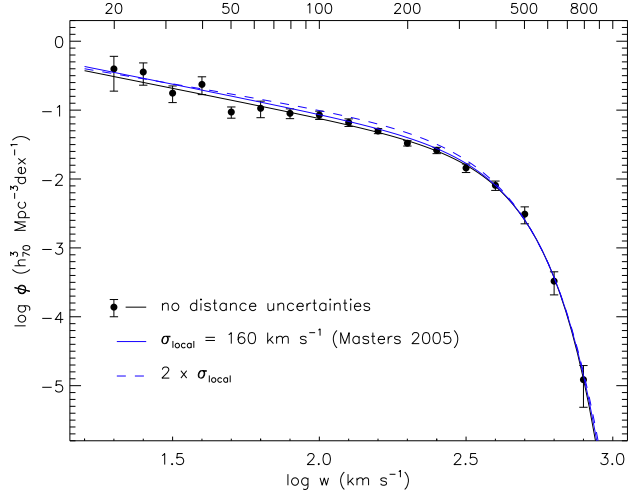}
\caption{\textit{Effect of distance uncertainties on the width function}:  Filled circles with error bars and black solid line as in Figure \ref{fig:wf}. The blue solid line corresponds to the result of adding  a random velocity dispersion of $\sigma_{local} = 160$ \kms \ \citep{2005PhDT.........2M} to the $\alpha$.40 galaxy distances. The dashed blue line corresponds to twice the fiducial velocity dispersion, $\sigma_{local} = 320$ \kms. Note the relative immunity of the WF against distance uncertainties. The main effect appears to be an overall increase in amplitude, while (in contrast to the case of the HIMF) no clear trend for a steepening of the low-end slope seems to exist (see \S\ref{subsec:eddington} for more details).}
\label{fig:eddington}
\end{figure}

The results are displayed graphically in Figure \ref{fig:eddington}, where the datapoints and solid line correspond to the original ALFALFA WF, and the blue solid and dotted lines correspond respectively to the results of the $\sigma_{local} = 160$ \kms \ and $\sigma_{local} = 320$ \kms \ simulation sets. The largest effect is an overall increase in the amplitude of the WF at intermediate widths, which is probably due to the net transport of sources towards lower HI masses and therefore larger values of $1/V_{eff}$. Unlike in the case of the HIMF, the low end slope $\alpha$ does not seem to be affected in any systematic way. We conclude that, apart from a mild increase in amplitude at intermediate widths, the WF is relatively insensitive to distance uncertainties due to galaxy peculiar motions.

\subsection{Cosmic Variance}
\label{subsec:cosmic}

\begin{figure}[p]
\includegraphics[scale=0.6]{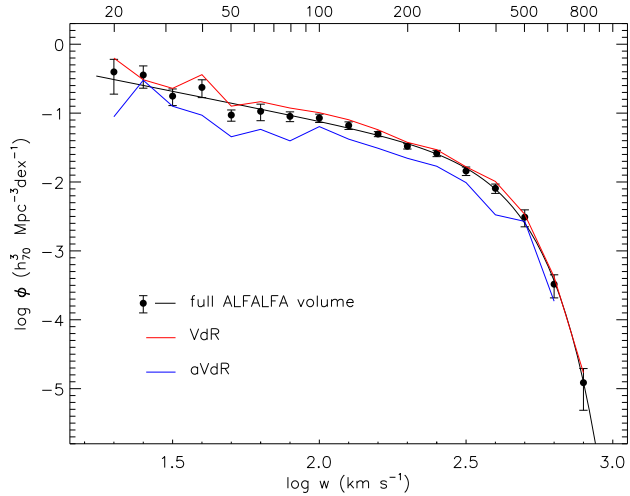}
\caption{\textit{Cosmic variance}: datapoints with error bars and black solid line as in Figure \ref{fig:wf}. The red and blue solid lines represent the WF in the Virgo direction Region (VdR: $07^h30^m < \alpha < 16^h30^m$, $4^\circ < \delta < 16^\circ$ and $24^\circ < \delta < 28^\circ$) and the anti-Virgo direction Region (aVdR: $22^h < \alpha < 03^h, \; 14^\circ < \delta < 16^\circ$ and $24^\circ < \delta <32^\circ$), respectively. The VdR is a locally overdense region while the aVdR is locally underdense, a fact that is reflected by the difference between the the two WFs at intermediate and low widths (see \S\ref{subsec:cosmic}).}
\label{fig:springfall}
\end{figure}

The WF presented in Figure \ref{fig:wf} aspires to represent the distribution in a cosmologically representative volume. The sensitivity of the ALFALFA survey allows $\sim w_\ast$ and broader galaxies to be detected throughout the full $\alpha.40$ volume ($V_{\alpha.40} \approx 3.1 \cdot 10^6 \;\; h_{70}^3 \, \mathrm{Mpc}^3$), which ensures a cosmologically fair sampling of the MW-sized galaxy population. On the other hand, low-width galaxies tend to be faint systems that can only be detected in smaller volumes. As a result, the low-width bins of the WF are subject to increased uncertainty caused by the deviation of the galaxy distribution from homogeneity on small scales, which is referred to as cosmic variance (see Figure \ref{fig:springfall} for a graphical illustration).

In order to quantitatively asses the effects of cosmic variance on the ALFALFA WF, we jackknife resample the $\alpha$.40 survey volume, by splitting it into 14 parts equally spaced in R.A. Then, we reevaluate the WF excluding each part in turn. The resulting scatter for each parameter, $x$, is given by  

\begin{equation}
\sigma_x^2 = \frac{N-1}{N} \: \sum(x-\bar{x})^2 \;,  \;\;\; N=14.
\label{eqn:scatter}
\end{equation}

\noindent
The scatter calculated by Eqn. \ref{eqn:scatter} would be equal to the purely statistical error if the survey volume were homogeneous, and so any excess noise results from the presence of inhomogeneities. The method described above provides a measurement of cosmic variance on linear scales smaller than those probed by the full survey, and hence yields a conservative estimate of the true uncertainty (cosmic variance generally increases with decreasing scale).

The full uncertainties on the fit parameters (including cosmic variance) are $\phi_\ast = 0.011 \, \pm \, 0.003 \; (0.002) \;\; h_{70}^3 \; $Mpc$^{-3}$dex$^{-1}$, $\log w_\ast = 2.58\,  \pm \, 0.04 \; (0.03)$, $\alpha = -0.85\, \pm \, 0.19 \; (0.10)$ and $\beta = 2.7 \, \pm \, 0.3 \; (0.3)$, where the term in parentheses represents the purely Poisson error reported in \S\ref{subsec:aawf}. Indeed, parameters $w_\ast$ and $\beta$, which dictate the shape of the WF at high widths, show a very modest increase in their uncertainty due to cosmic variance. On the other hand, the narrow end slope $\alpha$ is significantly affected, with cosmic variance contributing a large fraction of the full error.

\subsection{Beam confusion}

Beam confusion arises from the fact that the ALFA $3.3^\prime$ x $3.8^\prime$ beam occasionally produces blends of small galactic groups at moderate distances, especially when individual galaxies are poorly separated in redshift space. The qualitative effect of such blends is to transform two or more independent sources into a single HI profile of larger $w_{50}$ than each of its constituents. We do not attempt to quantify the effect of confusion bias, but we anticipate it to be more pronounced at the high-width end of the WF. This is because galaxies with $w \gtrsim 550$ \kms \ are preferentially found at large distances, where beam confusion is more severe. It is worth noting that this bias, even though present, cannot account for the discrepancy between the ALFALFA and HIPASS WFs at high widths, since the latter suffers from more confusion due to its larger beam size ($13^\prime$ FWHM).

\section{Comparison with Theory and Simulations}
\label{sec:theory}

The velocity function of halos in a CDM universe scales as $dn \propto v^{-4} \, dv$, where $v$ refers to the halo maximum rotational velocity. Even though a straightforward comparison of the CDM VF with the ALFALFA WF is not possible, such a steep scaling suggests a substantial discrepancy between the theoretical and observed distributions at low velocities. 
  
In order to make a meaningful comparison between the theoretical prediction and the ALFALFA measurement, it is necessary to take into account a number of important effects: 

\begin{enumerate}[i)]
 
\item DM halos exhibit significant substructure. High-resolution simulations have shown that several local density maxima (subhalos) develop within the virial radius of an underlying bound overdensity. As a result, massive halos ($M_{halo} \gtrsim 10^{13} \; h^{-1} \, M_\odot$) typically host groups or clusters of galaxies rather than a single astronomical object. In general, a one-to-one correspondence between simulated halos and visible galaxies is not always possible. 
\label{item:i}
\item The collapse of baryons to the central region of DM halos affects the galactic potential and leads to a modification of the true galactic rotation curves compared to the ones obtained in dissipationless DM simulations.
\label{item:ii}
\item The detectability of a galaxy in an HI survey depends on its atomic hydrogen content. Galaxies that are deficient in HI may be underrepresented in an HI-selected sample.
\label{item:iii}
\item The relationship between the maximum of the rotation curve of a galaxy and its HI velocity width is non-trivial. Apart from the obvious dependence on disk inclination, the measured width depends on the spatial distribution of atomic hydrogen in the galactic potential. In particular, HI disks do not always extend far enough to sample the asymptotic outer part of the galactic rotation curve, and may underestimate the mass of the host halo.  
\label{item:iv}

\end{enumerate}       

\noindent
It is, thus, necessary to populate the DM halos of an N-body simulation with modeled \textit{galaxies}, and compare this virtual sample against the ALFALFA measurement. Modeling of the atomic hydrogen content of the synthetic galaxies is particularly desirable, because it greatly facilitates the comparison between theoretical and observed distributions.

\begin{figure}[p]
\includegraphics[scale=0.6]{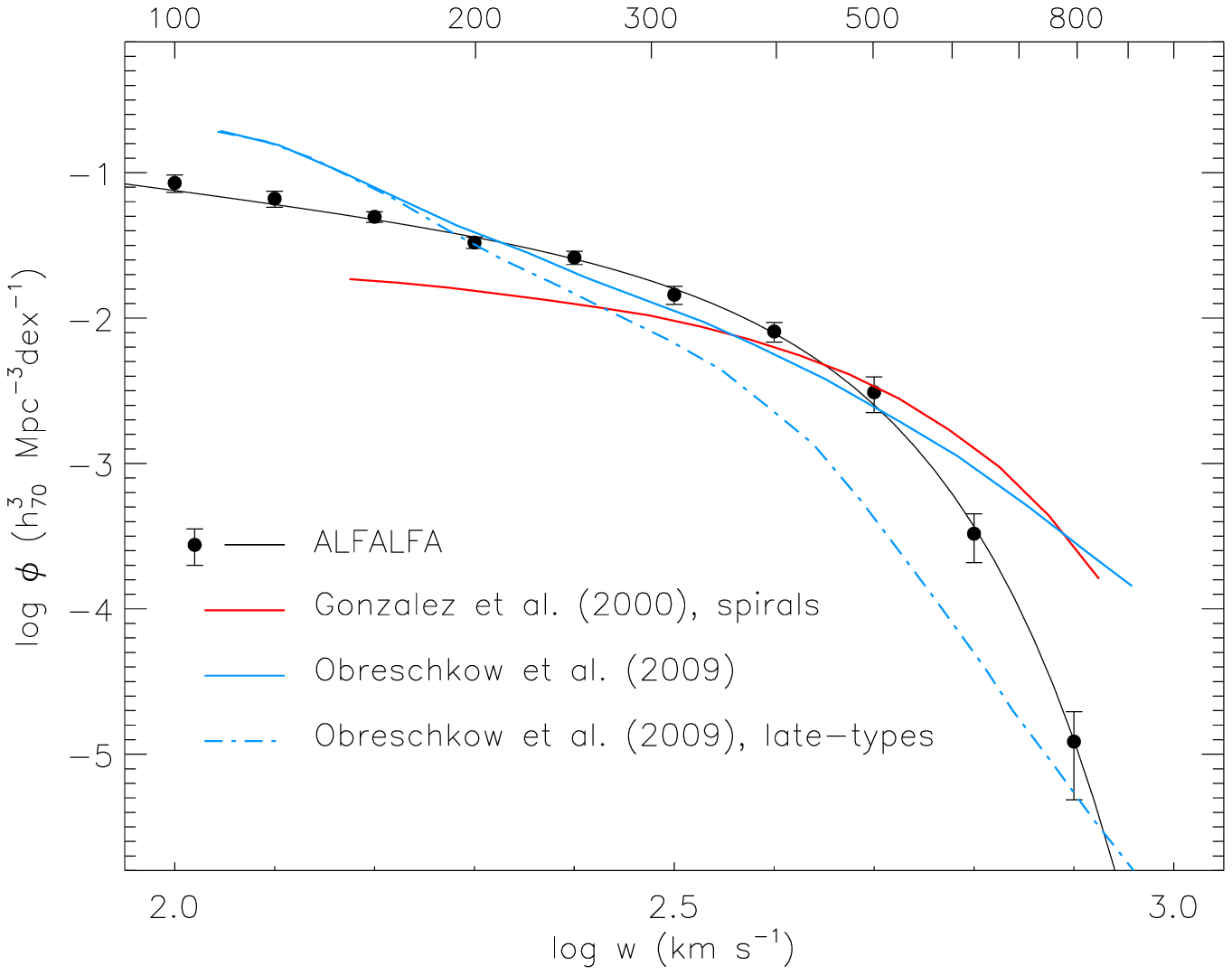}
\caption{Datapoints with errorbars and the black solid line represent the ALFALFA WF in the width range $w \geqslant 100$ \kms. The cyan solid line represents the \citet[O09]{2009ApJ...698.1467O} WF, derived from projecting their distribution of modeled HI linewidths ($w^{HI}_{50}$) for the synthetic galaxies in the semi-analytic catalog of \citet{2007MNRAS.375....2D}. The cyan dash-dotted line represents the subsample of the O09 galaxies classified as ``late-types'' according to their bulge-to-stellar mass ratios in the DeLucia catalog. The red solid line represents the projection of the indirect observational determination of the velocity function (VF) of spiral galaxies by \citet{2000ApJ...528..145G}. Their VF was obtained by combining the  observed luminosity function (LF) for spiral galaxies with the Tully-Fisher relationship.}
\label{fig:obreschkow}
\end{figure}

\citet[hereafter O09]{2009ApJ...698.1467O} have simulated the HI-line profiles for the galaxies in the \citet{2007MNRAS.375....2D} semi-analytic catalog, created by post-processing the Millennium N-body simulation \citep{2005Natur.435..629S}. Figure \ref{fig:obreschkow} displays the WF (cyan solid line) resulting from projecting their modeled edge-on linewidths, assuming random galaxy inclinations. The O09 WF is in fairly good agreement with the ALFALFA measurement, but fails to display an exponential cutoff at high widths and therefore predicts too many high-width galaxies. This issue has been also pointed out in Zw10, who argue that the disagreement is caused by the fact that the O09 catalog overestimates the HI masses of massive early-type galaxies. They found that restricting themselves to synthetic galaxies classified as late-types (based on their bulge-to-total stellar mass ratios in the DeLucia catalog) produced a much better fit to their data. However, Figure \ref{fig:obreschkow} suggests that applying the ``morphological'' cut of Zw10 results in too few galaxies at intermediate widths ($200$ \kms $< w < 600$ \kms). 

The red solid line in Figure \ref{fig:obreschkow} is the WF corresponding to an indirect observational estimate of the velocity distribution of spiral galaxies by \citet{2000ApJ...528..145G}. Their determination of the spiral galaxy VF was produced by combining the Southern Sky Redshift Survey $B$-band LF for spirals in conjunction with the \citet{1997ApJS..108..417Y} Tully-Fisher parameters in the $B_T$-band. This indirect method, based on galaxy scaling relations, is reliable only for relatively massive spirals ($v_{rot} > 70$ \kms) and suffers from numerous sources of uncertainty (e.g.  scatter in the TF relation, uncertainties related to the correction of the LF for extinction, bandpass conversion uncertainties, etc.).

\subsection{The CDM overabundance problem}
\label{subsec:overabundance}

CDM predictions start diverging from the observational results at low widths, and so the behavior of the theoretical WF for $w < 200$ \kms \ is of great importance. Unfortunately, the very interesting work of O09 is only reliable for $w \gtrsim 100$ \kms \ due to the limitations in the mass resolution of the Millennium simulation. We employ instead two recent high-resolution CDM simulations, that lack however modeling of the HI component of their virtual galaxy samples.     

\begin{figure}[p]
\includegraphics[scale=0.6]{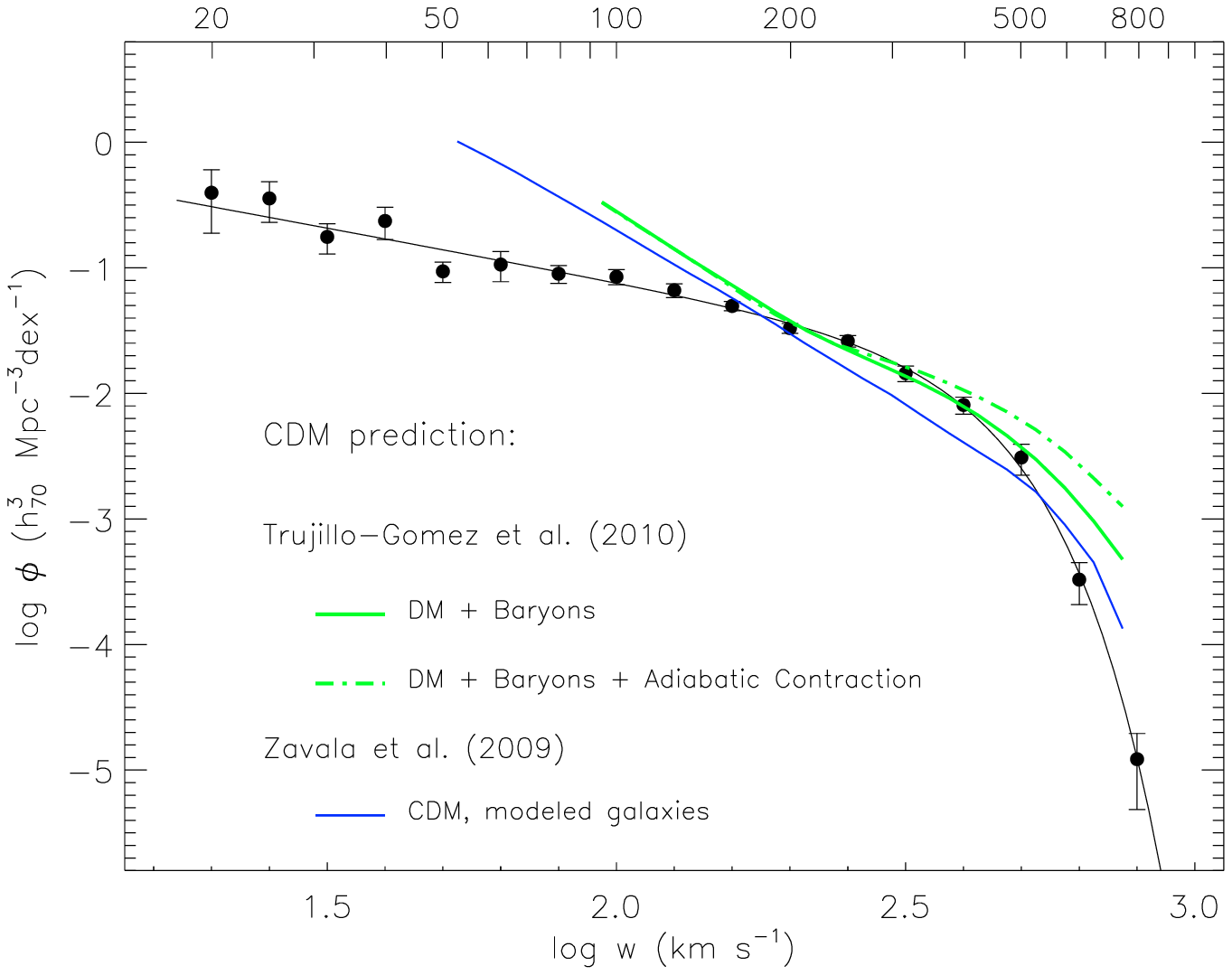}
\caption{\textit{The CDM overabundance problem}: datapoints with errorbars and black solid line represent the measured ALFALFA WF (same as in Figure \ref{fig:wf}). The green lines represent the WF of a sample of synthetic galaxies modeled by \citet[TG10]{2010arXiv1005.1289T}, which populate the halos in the Bolshoi CDM simulation \citep{2010arXiv1002.3660K}. Two models were considered by TG10, one where the gravitational potential of baryons is simply superimposed on the DM potential (solid line) and one where the subsequent adiabatic contraction of the DM halo is taken into account (dash-dotted line). The blue solid line represents the WF of a modeled galaxy population corresponding to the higher resolution CDM simulation of \citet[Za09]{2009ApJ...700.1779Z}. Note that both theoretical distributions predict a steeply rising low-width end, in stark contrast with the observational result. The discrepancy according to the Za09 result is a factor of $\sim$8 at $w = 50$ \kms, rising to a factor of $\sim 100$ when extrapolated to $w = 20$ \kms \ (see \S\ref{subsec:overabundance}).}
\label{fig:trujillo}
\end{figure}

Figure \ref{fig:trujillo} compares the ALFALFA measurement with the WF of the galaxy population corresponding to the Bolshoi simulation \citep{2010arXiv1002.3660K}, as modeled by \citet[hereafter TG10]{2010arXiv1005.1289T}. Each Bolshoi halo was assigned realistic stellar and cold gas masses, based on empirical relations. Subsequently, two models were considered, one where the gravitational potential of the baryons is simply superimposed on the DM potential (solid green line) and one where the DM halo adiabatically contracts in response to the presence of the baryons (dash-dotted green line). Note that TG10 define $v_{rot}$ as the value of the simulated rotation curve at a radius of 10 kpc. The authors argue that their modeling scheme and use of $v_{10kpc}$ provide a good approximation of the measured velocity for galaxies with both flat and rising rotation curves.

Also plotted in Figure \ref{fig:trujillo} is the WF of simulated galaxies based on the \citet[hereafter Za09]{2009ApJ...700.1779Z} constrained N-body simulation (blue solid line). Za09 perform a modest volume ($64 \: h^{-1}$ Mpc on a side) but very high-resolution ($v_{lim} = 24$ \kms) constrained simulation, designed to reproduce the large-scale structure of the local universe. Virtual galaxies are modeled according to the analytical results of \citet{1998MNRAS.295..319M}, assuming a disk-to-virial mass ratio of $f_{disk} \equiv M_{disk} / M_{vir} =0.03$ independent of halo size. Lastly, the maximum amplitude of the rotation curve ($v_{rot,max}$) for each galaxy is calculated, by combining the disk and DM halo contributions.    

Since neither model considers the distribution of the velocity field tracer (i.e. HI) in simulated galaxies, we convert rotational velocities into HI velocity widths by assuming the relationship 

\begin{equation}
w = 2 \: v_{rot} \:  \sin i + w_{eff} \;\; .
\label{eqn:vtow}
\end{equation} 

\noindent
Galaxies are assumed to be randomly oriented with respect to the line-of-sight ($\cos i$ is uniformly distributed in the $[0,1]$ interval), while $w_{eff}$ is a small ``effective'' term used  to reproduce the broadening effect of turbulence and non-circular motions on HI linewidths. The use of eqn. \ref{eqn:vtow} is \textit{only} justified if the HI disk is extended enough to sample the value of $v_{rot}$ adopted by the model under consideration (e.g. $v_{10kpc}$ for TG10 and $v_{rot,max}$ for Za09). We adopt the value $w_{eff} = 5$ \kms \ for the broadening term\footnote{The value of $w_{eff} =5$ \kms \, is derived empirically by \citet{2001A&A...370..765V}, based on a sample of 22 galaxies with flat or decreasing outer rotation curves.}, which is added linearly for galaxies with $v_{rot} > 50$ \kms \ and in quadrature for lower velocity galaxies.

Figure \ref{fig:trujillo} puts in evidence the marked departure of the theoretical distributions from the ALFALFA measurement at $w < 200$ \kms, which becomes more dramatic with decreasing width. According to the TG10 WF, the difference is approximately a factor of $\sim4$ at $w = 100$ \kms, exhibiting an increasing trend. The Za09 WF\footnote{In order to account for the fact that the Za09 sample resides in an overdense volume (within a radius of $20 \: h^{-1}$ Mpc from their simulated ``Local Group''), we lower the normalization of their WF by a factor of 2, as suggested in their \S 4.3.}, implies a difference of a factor of $\sim 8$ at the lowest width where the simulation is complete ($w \approx 50$ \kms), and displays a much steeper low-width slope than the ALFALFA measurement. An extrapolation of the Za09 WF to the ALFALFA width limit ($w = 20$ \kms), would result in a discrepancy of a factor of $\sim 100$.

\subsection{Is CDM viable?}
\label{subsec:solutions}

The ALFALFA measurement of the WF confirms the results of the HIPASS survey \citep{2010MNRAS.403.1969Z}, which obtained its WF at lower sensitivity and velocity resolution. This fact excludes the possibility that the CDM overabundance problem is an artifact of the limited performance characteristics of past blind 21 cm surveys. The reason for the observed discrepancy can be therefore most likely attributed to one of the two following factors:    

\begin{enumerate}[1.]

\item The inaccuracy of standard CDM simulations, presumably due to the inadequacy of the assumed DM model.
 
\item The improper comparison of simulated halos with observed galaxies. This could be due either to
\begin{enumerate}[a)]  
\item the inadequate modeling of the baryonic counterparts hosted by DM halos, which leads to wrong predictions for maximum rotational velocities, or
\item the incorrect interpretation of inclination-corrected HI linewidths as maximum rotational velocities.
\end{enumerate}  

\end{enumerate}

In what follows, we will consider these possibilities in more detail and argue about their prospects as solutions of the CDM overabundance problem.

Most large-scale simulations of cosmic structure conform to the standard $\Lambda$CDM cosmological model. In particular, they assume that all dark matter is cold (i.e. has negligible free-streaming length), non self-interacting and stable (i.e. non-decaying). These properties are appropriate for a universe where dark matter consists of stable weakly interacting massive particles (WIMPs). WIMPs are currently the favored DM particle candidate, and are expected to have masses in the GeV-TeV range and weak scale self-interaction cross-sections, justifying the DM attributes most commonly assumed in cosmological N-body simulations. 

\begin{figure}[p]
\includegraphics[scale=0.6]{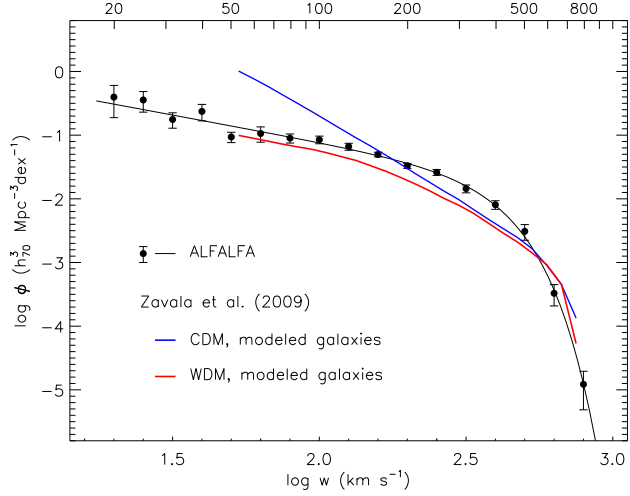}
\caption{Data points with error bars and black solid line represent the measured ALFALFA WF (same as in Figure \ref{fig:wf}). The blue solid line represents the WF of a modeled galaxy population based on the high resolution CDM simulation of \citet[Za09]{2009ApJ...700.1779Z} (same as in Figure \ref{fig:trujillo}). The red solid line represents the WF corresponding to a second run of the Za09 simulation assuming a 1 keV WDM particle (both simulations employ the same scheme to populate halos with synthetic galaxies). The WDM WF displays a shallow low-width slope due to the suppressed formation of structure at small scales, and is in much better agreement with the ALFALFA measurement.}
\label{fig:zavala}
\end{figure}

However, the picture changes considerably if DM is composed of relatively light ($\sim$ keV) particles, in which case it is referred to as warm dark matter (WDM). Structure on large scales would be the same as in a CDM universe, but on small scales halo formation would be heavily suppressed due to the non-negligible free-streaming length of the light WDM particle. Za09 have considered this alternative scenario, and carried out a second run of their very high-resolution simulation assuming a 1 keV WDM particle. They subsequently populate their halos with synthetic galaxies, employing the same modeling scheme as in their CDM run (\S\ref{subsec:overabundance}). The result is shown by the red solid line in Figure \ref{fig:zavala}, superposed on the ALFALFA WF (datapoints with errorbars and black solid line) and the result of their CDM run (blue solid line). 

Strikingly, the synthetic WF in the WDM case exhibits a shallow slope at the low-width end, in good agreement with the slope measured by ALFALFA. Such a shallow slope results from the suppressed production of low-mass halos in a WDM universe, which directly translates into a lower abundance of low-width visible galaxies. WDM could therefore provide a simple and elegant solution of the overabundance problem.

Despite its appeal in this specific context, the general prospects of WDM also depend on its overall viability as the dominant constituent of non-baryonic matter in the universe. A number of theoretical microscopic models for WDM have been proposed, most commonly involving sterile neutrinos \citep{1994PhRvL..72...17D, 2003PhRvD..68j3002F, 2005PhLB..631..151A, 2009PhR...481....1K}. Constraints on the particle's mass can be placed by astrophysical and cosmological considerations. In particular, Ly$\alpha$ forest data places lower limits on the neutrino mass (a lighter particle generally results in suppression of power at larger scales), while X-ray observations can place upper mass limits (radiative decay into X-ray photons generally becomes more efficient at higher masses). The limits on the neutrino mass imposed by these observational constraints depend on the assumed neutrino production mechanism. \citet{2006PhRvD..74b3527A} find that non-resonantly produced neutrinos are ruled out, using a compilation of Ly$\alpha$ forest and X-ray data (see references therein). \citet{2009PhRvL.102t1304B} have considered sterile neutrino production in the context of the $\nu$MSM (Minimal Standard Model + 3 sterile neutrinos) and argue that neutrinos with $m_{sn} >$ 2 keV are viable.

The second class of potential solutions attribute the disagreement between theory and observation to the process used to translate the output of simulations into actual galaxies. In particular, a number of important effects need to be taken into account  (identified as items \ref{item:i}-\ref{item:iv} in Section \ref{sec:theory}) to ensure a successful comparison of simulated halos with observed galactic samples.

Both theoretical works presented in \S\ref{subsec:overabundance} address issues \ref{item:i} and \ref{item:ii}. For example, Za09 set an explicit limit on the mass of halos hosting individual galaxies at $M_{vir} = 10^{13} \; h^{-1} \: M_\odot$. The influence of baryons on the shape of galactic rotation curves is also taken into account by both works, albeit using slightly different prescriptions and definitions of galaxy rotational velocity. Despite the use of numerous simplifying assumptions by TG10 (e.g. all baryons within 10kpc) and Za09 (e.g. fixed disk-to-virial mass ratio for all galaxies) their theoretical WFs  are in fair agreement with the ALFALFA measurement at intermediate widths ($200$ \kms $< w < 500$ \kms).

The last two issues are related specifically to the atomic hydrogen content of galaxies, which is not modeled by either TG10 or Za09. Specifically, issue \ref{item:iii} concerns the detectability of a galaxy in a 21cm survey. In principle, there exists the possibility that most of the low-mass halos predicted by CDM cosmology correspond to HI-devoid, dwarf spheroidal galaxies. In reality, a solution involving a multitude of isolated early-type dwarf systems seems rather unlikely. Direct observations \citep{2002ApJ...581.1019G, 2002A&A...390..863S, 2005A&A...442..137N}, as well as other empirical arguments, suggest that the HI-to-stellar mass ratio grows with decreasing mass for galaxies in the field. HI surveys should thus have an advantage, rather than a disadvantage, at detecting the baryonic counterparts hosted by low-mass DM halos. In addition, optical surveys suggest that isolated early-type dwarfs in medium/low density environments are relatively rare \citep{2004AJ....127.2031K}. A second issue relates to the fact that satellite galaxies may be underrepresented in the $\alpha$.40 sample, since they are generally redder (and have presumably lower gas fractions) than central galaxies of the same luminosity \citep[e.g.][]{2008MNRAS.389.1619F}. This bias could result in a $\lesssim  30\%$ underestimate of the abundance of low-width galaxies by ALFALFA  \citep[e.g][]{2008ApJ...676..248Y,2010arXiv1002.3660K}, not nearly enough to explain the observed discrepancies.

Issue \ref{item:iv} regards the size and detailed spatial distribution of the atomic hydrogen component in galaxies, which determines the way in which its rotation curve is converted into an HI velocity width. In particular, $w_{HI}$ is a fair tracer of the maximum rotational velocity, only if the HI disk is extended enough to reach the flat (or decreasing) part of the galactic rotation curve.     
The use of Eqn. \ref{eqn:vtow} in the derivation of the theoretical WFs implicitly assumes this situation to be true; observationally however, this is often times not the case. For example, the \citet{2006ApJ...640..751C} set of template rotation curves, puts in evidence the fact that lower rotational velocity galaxies tend to have steeper outer rotation curves (see their Figure 1 \& 4). The dwarf galaxy samples of \citet{2005AJ....129.2119S} and \citet{2009A&A...493..871S}, suggest that the effect becomes more dramatic at the lowest velocities (see Figure 3 \& Figure 4 in the respective references).

This systematic trend for lower velocity galaxies to host less extended HI disks can be understood in terms of the expected baryon depletion of low-mass halos. Results from N-body + hydrodynamics simulations \citep[e.g.][]{2006MNRAS.371..401H, 2010AdAst2010E..33R} indicate that halos with mass below some critical value lose a significant fraction of their cosmic share of baryonic matter, due to environmental and internal feedback processes. In particular, UV heating of the intergalactic medium (IGM) after reionization is believed to lead to substantial gas removal from low-mass halos ($v_{rot} \lesssim 20$ - $30$ \kms, corresponding to $M_{vir} \lesssim 10^{9}$ - $10^{9.5} \; h^{-1} \: M_\odot$). Internal feedback processes such as supernova winds may also be important, but their efficacy is strongly model dependent.

The above considerations could lead to a solution of the overabundance problem that would not require a modification of the extremely successful $\Lambda$CDM paradigm. In simple terms, the overabundance problem would be the result of the inability of HI to trace the maximum halo rotational velocity of low-mass systems, which leads to a severe underestimate of their true mass. The same argument has been identified as a possible solution of the ``mini-void size'' problem by \citet{2009MNRAS.395.1915T}, while a similar effect has been proposed by \citet{2008ApJ...673..226P} as a solution to the ``missing satellites'' problem. 


\section{The $v_{rot}$ - $v_{halo}$ relation in a CDM universe}
\label{sec:relation}       
 
Assuming the CDM model to be correct, we can statistically infer the $v_{rot}$ - $v_{halo}$ relationship needed to reproduce the observational galaxy VF. This can be done by abundance matching, a statistical procedure which assumes the existence of a one-to-one relationship between galaxy and halo circular velocities, $v_{rot} = f(v_{halo})$. It follows that the space density of halos with circular velocities larger than a given value, $V$, should be equal to the space density of galaxies with rotational velocities larger than the value dictated by the relationship, $n(v_{halo} > V) = n(v_{rot} > f(V))$. 

Obtaining an observational velocity distribution from the ALFALFA measurement is not straightforward. Firstly, the ALFALFA measurement regards galaxy velocity widths (uncorrected for inclination) and not intrinsic rotational velocities; secondly, the ALFALFA survey is biased against HI-poor massive ellipticals that dominate the counts at high velocities. 

\begin{figure}[p]
\includegraphics[scale=0.6]{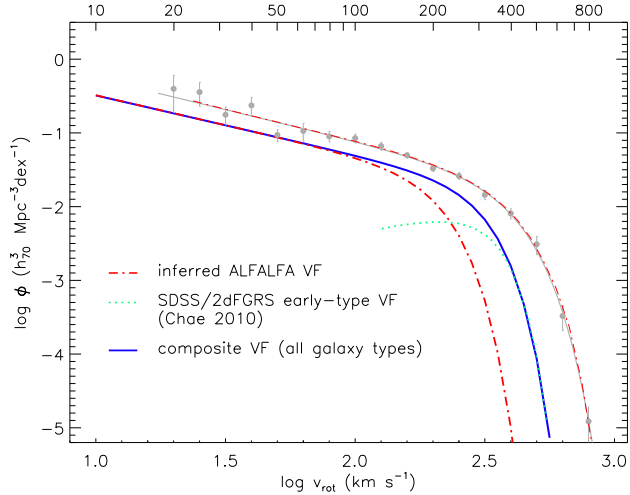}
\caption{The thick red dash-dotted line represents the velocity function of late-type galaxies (assumed to follow a modified Schechter distribution) that best reproduces the measured ALFALFA WF (light gray data points and solid line) upon projection (thin red dash-dotted line). The green dotted line represents the velocity function of early-type galaxies determined by \citet{2010MNRAS.402.2031C} using SDSS and 2dFGRS data. The blue solid line is a modified Schechter interpolation of the two VFs which represents a velocity function valid for all morphological types. The modified Schechter parameters for the interpolated distribution are $\phi_\ast = 8.7 \cdot 10^{-3} \; h_{70}^3 \: \mathrm{Mpc}^{-3}$, $\log v_\ast = 2.49$, $\alpha = -0.81$ and $\beta = 3.35$ (see Section \ref{sec:relation} for more details).}     
\label{fig:vf}
\end{figure}

We address the first issue by searching for the velocity distribution that best reproduces the ALFALFA WF, upon projection using Eqn. \ref{eqn:vtow}. We assume that the distribution follows a modified Schechter function of the form

\begin{equation}
\phi(v) = \frac{dn}{d\log v} = \ln(10) \: \phi_\ast \left(\frac{v}{v_\ast}\right)^\alpha e^{-(\frac{v}{v_\ast})^\beta} \;\;,
\end{equation}

\noindent
and that it corresponds to the VF of HI-rich, late-type galaxies. The set of parameters that provide the best match is identified visually, and corresponds to the values $\phi_\ast = 1.2 \cdot 10^{-2}\;\; h_{70}^3 \: \mathrm{Mpc}^{-3}$, $\log v_\ast = 2.32$, $\alpha = -0.81$ and $\beta = 3.1$ (thick red dash-dotted line in Figure \ref{fig:vf}). In order to address the second issue (i.e. obtain a VF valid for all morphological types), we use the results of \citet{2010MNRAS.402.2031C}, who studied the velocity dispersion function (VDF) of early-type galaxies in the SDSS and 2dFGRS surveys. Velocity dispersions can be transformed into rotational velocities by assuming an isothermal mass profile, in which case $v_{rot} = \sqrt{2}\sigma$. We adopt the average of the 2dFGRS and SDSS velocity distributions as a representative VF for early-type galaxies, which we plot as the green dotted line in Figure \ref{fig:vf}.

\begin{figure}[p]
\includegraphics[scale=0.6]{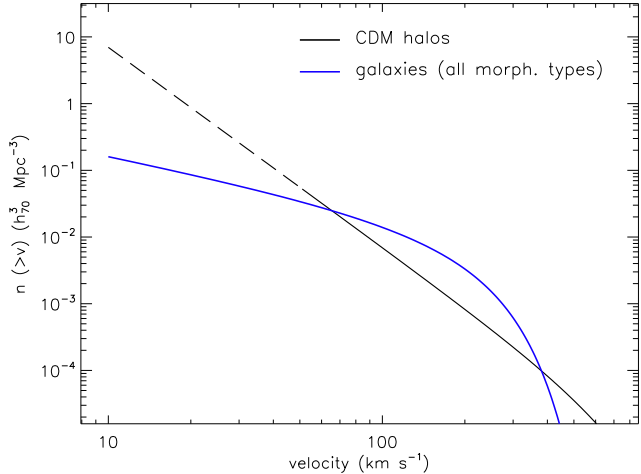}
\caption{The velocity function of halos (black line) and galaxies (blue line), expressed as a cumulative distribution. The former distribution corresponds to the number density of halos (including subhalos) in the Bolshoi CDM simulation, as a function of their maximum rotational velocity at the present epoch ($v_{halo}$). Note that the Bolshoi simulation is complete only down to $v_{halo} = 50$ \kms, but a power-law extrapolation to lower velocities (black dashed line) is expected to hold. The latter distribution represents the VF of all galaxy types, as a function of their observed rotational velocity (same as blue line in Figure \ref{fig:vf}, see discussion is Section \ref{sec:relation}).}
\label{fig:cumvfs}
\end{figure}

We interpolate the two distributions using a single modified Schechter function with parameters $\phi_\ast = 8.7 \cdot 10^{-3} \;\; h_{70}^3 \: \mathrm{Mpc}^{-3}$, $\log v_\ast = 2.49$, $\alpha = -0.81$ and $\beta = 3.35$. The interpolated distribution (blue solid line in Figure \ref{fig:vf} \& Figure \ref{fig:cumvfs}) represents a composite galactic VF valid for all morphological types. Even though we do not formally measure errors for the derived distribution, we list below some important sources of uncertainty. Firstly, the statistical uncertainty on the parameters of the late-type VF should be at least on the order of the errors reported in \S\ref{subsec:aawf}; the parameters of the composite VF should be expected to carry larger errors, since the determination of the interpolating distribution is subjective to some extent. More importantly though, there are a number of sources of uncertainty related to galactic physics. For example, the assumption of isothermality of early-type galaxies is not expected to hold in detail \citep[e.g][]{2010MNRAS.407....2D}, which would affect the high-velocity end of the composite VF. Moreover, the low velocity slope, $\alpha$, depends partly on the value of $w_{eff}$ employed in Eqn. \ref{eqn:vtow}; the value adopted here ($w_{eff} = 5$ \kms) has been empirically determined from a sample of relatively massive spirals \citep{2001A&A...370..765V}, and does not have to be the same for galaxies populating the low-velocity end of the VF. Also, as mentioned in \S \ref{subsec:solutions}, the inferred VF may be underestimating the true abundance of low-width galaxies by $\lesssim 30$ \%, since ALFALFA is likely to miss some fraction of the satellites of massive spiral galaxies.

Next, we obtain the theoretical CDM VF from the Bolshoi simulation\footnote{The Bolshoi simulation is run for the set of cosmological parameters $h=0.70$, $\Omega_m =0.27$, $\sigma_8 =0.82$, $n = 0.95$.} \citep{2010arXiv1002.3660K}. In particular, we use the distribution of maximum halo rotational velocity, $v_{halo}$, of all simulated halos (including subhalos) at the present epoch, which is shown as the black solid line in Figure \ref{fig:cumvfs}. Note that the simulation is run for the total matter density of the universe ($\Omega_m = \Omega_{DM} + \Omega_{bar} = 0.27$), but both DM and baryons are treated as dissipationless components. Also note that the simulation is complete only down to $v_{halo} = 50$ \kms, and a power-law extrapolation is used at lower velocities \citep[which is however expected to hold, see for example \S4.2 in][]{2009ApJ...700.1779Z}.

\begin{figure}[p]
\includegraphics[scale=0.6]{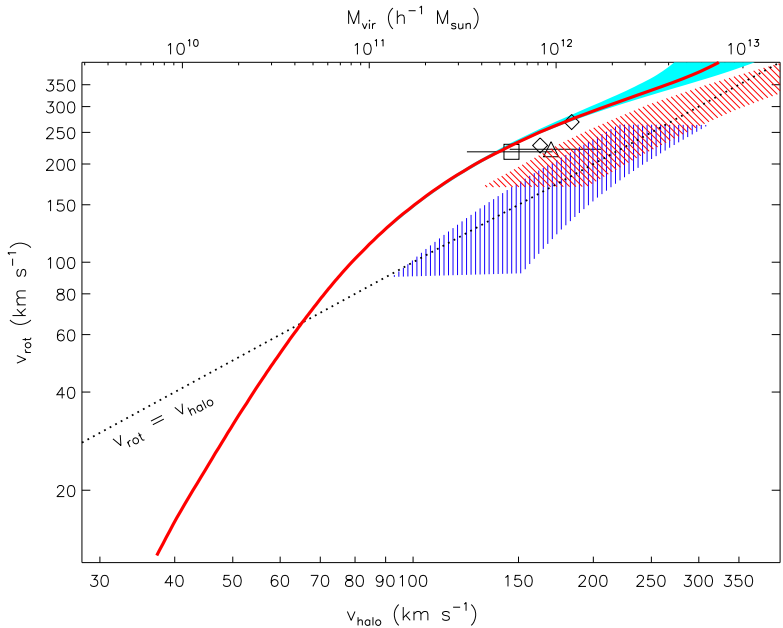}
\caption{$v_{rot}$ - $v_{halo}$ \textit{relation in a CDM universe}: the red solid line corresponds to the relationship between the rotational velocity of galaxies measured observationally ($v_{rot}$) and the maximum rotational velocity of the corresponding CDM halo ($v_{halo}$). The relation was obtained by the abundance matching of the velocity distribution of halos in the Bolshoi CDM simulation with the velocity distribution of galaxies inferred from ALFALFA and SDSS/2dFGRS data (see Figure \ref{fig:vf} \& \ref{fig:cumvfs}). We have assumed that halos with $v_{halo} > 360$ \kms \ ($M_{vir} \gtrsim 10^{13} \; h^{-1} \, M_\odot$) do not host individual galaxies, but rather groups of galaxies. The cyan shaded area corresponds to different mass cutoffs, ranging from $v_{halo,max} = 290$ \kms \ ({\it upper boundary}) to $v_{halo,max} = 440$ \kms ({\it lower boundary}). The blue and red hatched areas correspond to the $2\sigma$ error regions for late- and early-type galaxies respectively, according to \citet{2010MNRAS.407....2D}. Their measurement was based on a compilation of weak lensing and satellite kinematics measurements of galaxy dynamical masses (see Section \ref{sec:relation} for more details). The symbols correspond to the values estimated for the MW and M31 based on dynamical models \citep[\textit{diamonds}]{2002ApJ...573..597K}, and for the MW based on the kinematics of high velocity stars \citep[\textit{triangle with $2\sigma$ errorbars}]{2007MNRAS.379..755S} and blue horizontal branch stars \citep[\textit{box with $2\sigma$ errorbars}]{2008ApJ...684.1143X}. }

\label{fig:relation}
\end{figure}

The red thick line in Figure \ref{fig:relation} represents the $v_{rot}$ - $v_{halo}$ relation obtained by matching the CDM and galactic velocity distributions (values listed in Table \ref{tab}). We have assumed that halos with $v_{halo} > 360$ \kms \ ($M_{vir} \gtrsim 10^{13} \; h^{-1} \: M_\odot$) do not host individual galaxies but rather groups of galaxies, and are hence excluded from the matching process. The cyan shaded region corresponds to different values for this mass cutoff, ranging from $v_{halo,max}  = 290$ \kms  \ ($M_{vir} \approx 5 \cdot 10^{12} \; h^{-1} \: M_\odot$, \textit{upper boundary}) to $v_{halo,max} = 440$ \kms \ ($M_{vir} \approx 2 \cdot 10^{13} \; h^{-1} \: M_\odot$, \textit{lower boundary}). The uncertainty in the value of $v_{halo,max}$ mentioned above is the only source of error considered explicitly here. There are, however, additional uncertainties involved in the determination of the presented relationship. For example, no scatter in the $v_{rot} = f(v_{halo}) $ relation was considered in the abundance matching process. Also, no corrections to $v_{rot}$ for pressure support have been made in this work, even though gas thermal velocities in low mass galaxies can be comparable with their rotational velocities.

Figure \ref{fig:relation} shows that $v_{rot}$ follows an approximately linear relationship with $v_{halo}$ only for intermediate-mass halos (120 \kms \ $\lesssim v_{halo} \lesssim 170$ \kms). In this range, $v_{rot} \approx 1.5 \: v_{halo}$, in fair agreement with the values estimated for the MW and M31 from dynamical models \citep[{\it diamonds}]{2002ApJ...573..597K} and from the kinematics of MW high velocity stars \citep[{\it triangle}]{2007MNRAS.379..755S} and blue horizontal branch stars \citep[{\it box}]{2008ApJ...684.1143X}. However, the $v_{rot} / v_{halo}$ ratios obtained here are significantly larger than the average values inferred by \citet{2010MNRAS.407....2D} from a compilation of weak lensing and satellite kinematics datasets. Note though that their results are expressed in terms of a $v_{opt}$ - $v_{200}$ relation, where $v_{opt}$ is defined as the measured rotational velocity at 2.2 \textit{I}-band disk scalelengths for late-type galaxies and $1.65\sigma$ for early-type galaxies, and $v_{200}$ refers to the virial velocity of the halo at an enclosed overdensity of 200 times the critical density. In order to display their results in Figure\ref{fig:relation}, (red \& blue hatched regions) we have transformed virial velocities into halo maximum rotational velocities assuming average halo concentrations \citep{2008MNRAS.391.1940M}.  
 
The most important result of Figure \ref{fig:relation} concerns the low halo velocity regime. In  particular the relationship steepens continuously as we move to lower halo velocities, assuming a power-law behavior of the form $v_{rot} \propto v_{halo}^3$ at $v_{halo} \lesssim 50$ \kms. As a result, the true mass of low-velocity halos is systematically underestimated when measured by the inclination-corrected HI linewidth of the hosted galaxy; the underestimate can reach a factor of $\sim 2.5$ for $v_{halo} \approx 40$ \kms. Testing the low-velocity end of the $v_{rot}$ - $v_{halo}$ relation would require a sample of low-mass galaxies with directly measured dynamical masses, e.g. through weak lensing or satellite kinematics. However, some indirect observational support could come from a rich sample of HI interferometric maps of dwarf galaxies: a gradual transition from mostly flat to mostly rising rotation curves at $v_{rot} \approx 110$ \kms, would be required to explain the steepening of the relation at low velocities. Ultimately, observational verification of the presented relationship at low velocities would provide a check of the validity of the CDM model.

\section{Conclusions}
\label{sec:conclusions}       

We have measured the velocity width function (WF) of HI-bearing galaxies, based on a sample of 10,744 extragalactic sources detected in $\sim$40\% of the final ALFALFA survey area. The ALFALFA measurement extends to widths (uncorrected for inclination) as low as $w = 20$ \kms, and results in a robust measurement of the low-width logarithmic slope of $\alpha = -0.85 \,  \pm  \, 0.19$ ($1\sigma$ statistical error including the effect of cosmic variance). This result suggests a significant incompatibility of the observational distribution with the much steeper distribution expected in a CDM universe. 

We compare the ALFALFA result with the WFs of two modeled galaxy populations, one populating the Bolshoi CDM simulation halos \citep{2010arXiv1005.1289T} and the other populating the halos of the very-high-resolution CDM simulation of \citet{2009ApJ...700.1779Z}. Indeed, the simulated WFs start diverging from the ALFALFA measurement at widths $w \lesssim 200$ \kms. The difference in abundance is a factor of $\sim 8$ at $w = 50$ \kms \ (which corresponds to the resolution limit of the Za09 simulation), and implies a difference of a factor of $\sim 100$ when extrapolated to the ALFALFA low-width limit ($w = 20$ \kms). This discrepancy is closely related to a number of other observational challenges to CDM (e.g. ``missing satellites problem'', ``mini-void size problem'', etc.), which we collectively refer to as the \textit{CDM overabundance problem}.  

We further identify the two most promising solutions to the problem: the first involves the suppression of low-mass halo formation, which is best accomplished by assuming a $\sim$keV WDM particle; the second solution does not require a modification of the extremely successful CDM  model, and relies on the fact that HI disks in dwarf galaxies are frequently not extended enough to probe the full amplitude of the galactic rotation curve. The latter solution, supported by currently limited observational evidence, implies that galaxy rotational velocities derived from inclination-corrected HI linewidths ($v_{rot}$) systematically underestimate the maximum rotational velocity of their host DM halo ($v_{halo}$), below $v_{rot} \approx 110$ \kms. 

We furthermore use an abundance matching procedure to statistically infer the $v_{rot}$ - $v_{halo}$ relationship needed to reconciliate the CDM and galactic velocity distributions. We find that for MW-sized galaxies $v_{rot} \approx 1.5 \: v_{halo}$, while at low velocities $v_{rot}$ underestimates significantly the true maximum rotational velocity of the host halo.

Determining the correct solution to the CDM overabundance problem rests both on the general prospects of WDM as a viable dark matter model, as well as on observational verification of the $v_{rot}$ - $v_{halo}$ relationship predicted for CDM. The latter goal could be best accomplished through a rich sample of low-mass galaxies with directly measured dynamical masses.

\acknowledgements
The authors would like to acknowledge the work of the entire ALFALFA collaboration team in observing, flagging, and extracting the catalog of galaxies used in this work. This work was supported by NSF grant AST-0607007 and by grants from the Brinson Foundation. We would also like to thank an anonymous referee for their careful reading and their very helpful comments and suggestions.

\bibliographystyle{apj}
\bibliography{myreferences}

\begin{deluxetable}{cccc}
\tablewidth{0pt}
\tabletypesize{\tiny}

\tablecaption{\, The $v_{rot}$ - $v_{halo}$ relationship in a CDM universe\label{tab}}

\tablehead{\colhead{$v_{halo}$ [\kms]} & \colhead{$v_{rot}$ [\kms]}  &
\colhead{$v_{rot}$ [\kms]} & \colhead{$v_{rot}$ [\kms]}\\
\colhead{}&  \colhead{\scriptsize{($v_{halo,max} = 360$ \kms)}}  & \colhead{\scriptsize{($v_{halo,max} = 290$ \kms)}} & \colhead{\scriptsize{($v_{halo,max} = 440$ \kms)}}}

\startdata
40 & 16 & 16 & 16  \\
45 & 23 & 23 & 23 \\
50 & 32 & 32 & 32 \\
55 & 42 & 42 & 42 \\
60 & 53 & 53 & 53 \\
70 & 77 & 78 & 77 \\
80 & 102 & 103 & 102 \\
90 & 127 & 127 & 125 \\
100 & 149 & 150 & 147 \\
120 & 188 & 190 & 185 \\
140 & 219 & 223 & 218 \\
160 & 247 & 252 & 244 \\
180 & 270 & 278 & 267 \\
200 & 291 & 303 & 286 \\
220 & 310 & 327 & 303 \\
240 & 328 & 353 & 318 \\
260 & 345 & 383 & 333 \\
300 & 383 & \nodata & 360 \\
340 & 431 & \nodata & 387 \\
380 & \nodata & \nodata & 416 \\
420 & \nodata & \nodata & 449

\enddata


\end{deluxetable}

\end{document}